\documentstyle[12pt,aaspp4,tighten]{article}
 
\def \ref {\noindent\hangindent=1.0in\hangafter=1}
\def \cl {\centerline}

\def\ltsima{$\; \buildrel < \over \sim \;$}
\def\simlt{\lower.5ex\hbox{\ltsima}} 
\def\gtsima{$\; \buildrel > \over \sim \;$}
\def\simgt{\lower.5ex\hbox{\gtsima}} 

\begin{document}
 
\title{Hubble Space Telescope Imaging of the Optical 
Transient Associated with GRB970508}

\author{E. Pian\altaffilmark{1}, 
A. S. Fruchter\altaffilmark{2},
L. E. Bergeron\altaffilmark{2},
S. E. Thorsett\altaffilmark{3},
F. Frontera\altaffilmark{1,4},
M. Tavani\altaffilmark{5,6},}

\author{E. Costa\altaffilmark{7},
M. Feroci\altaffilmark{7},
J. Halpern\altaffilmark{5},
R. A. Lucas\altaffilmark{2},
L. Nicastro\altaffilmark{1},
E. Palazzi\altaffilmark{1},
L. Piro\altaffilmark{7},}

\author{
W. Sparks\altaffilmark{2},
A. J. Castro-Tirado\altaffilmark{8},
T. Gull\altaffilmark{9},
K. Hurley\altaffilmark{10},
H. Pedersen\altaffilmark{11}
}

\altaffiltext{1}{Istituto di Tecnologie e Studio delle Radiazioni 
Extraterrestri, C.N.R., Via Gobetti 101, I-40129 Bologna, Italy}

\altaffiltext{2}{Space Telescope Science Institute, 3700 San Martin Drive,
Baltimore, MD 21218, USA}

\altaffiltext{3}{Joseph Henry Labs. and Dept. of Physics, Princeton 
University, Princeton, NJ 08544, USA}

\altaffiltext{4}{Dip. Fisica, Universit\`a di Ferrara, Via Paradiso 12, 
I-44100 Ferrara, Italy} 

\altaffiltext{5}{Columbia Astrophysics Laboratory, Columbia University, 
New York, NY 10027, USA}

\altaffiltext{6}{Istituto di Fisica Cosmica e Tecnologie Relative, C.N.R., 
Via Bassini 15, I-20133 Milano, Italy}

\altaffiltext{7}{Istituto di Astrofisica Spaziale, C.N.R., Via E. Fermi 21, 
I-00044 Frascati, Italy}

\altaffiltext{8}{Laboratorio de Astrof\'{\i}sica Espacial y F\'{\i}sica 
Fundamental, INTA, P.O. Box 50727, 28080 Madrid, Spain}

\altaffiltext{9}{NASA/ Goddard Space Flight Center, Greenbelt, MD 20071, USA}

\altaffiltext{10}{University of California Space Sciences Laboratory, 
Berkeley, CA 94720, USA}

\altaffiltext{11}{Copenhagen University Observatory, Juliane Maries Vej 
30, DK 2100 Copenhagen, Denmark}

\begin{abstract}

We report on Hubble Space Telescope (HST) observations of the optical
transient (OT) discovered in the error box of the gamma-ray burst
GRB970508.  The object was imaged on 1997 
June 2 with the Space Telescope 
Imaging Spectrograph (STIS) and Near-Infrared Camera and Multi-Object 
Spectrometer (NICMOS).  The observations reveal a point-like source with 
R = $23.1 \pm 0.2$ and H = $20.6 \pm 0.3$, in agreement with the power-law 
temporal decay seen in ground-based monitoring.  Unlike the case of 
GRB970228, no nebulosity is detected surrounding the OT of GRB970508.  
We set very conservative upper limits of R $\sim 24.5$ and H $\sim 22.2$ 
on the brightness of any underlying extended source.  If this subtends a 
substantial fraction of an arcsecond, then the R band 
limit is $\sim$25.5.  In combination with Keck spectra that show Mg I 
absorption and [O II] emission at a redshift of $z=0.835$, our observations 
suggest that the OT is located in a star-forming galaxy with total 
luminosity one order of magnitude lower than the knee of the galaxy 
luminosity function, $L^*$.  Such galaxies are now thought to harbor the 
majority of star formation at $z\sim1$; therefore, these observations may 
provide support for a link between GRBs and star formation.
   
\end{abstract}
 
 
\section{Introduction}
 
Gamma-Ray Bursts (GRBs) were discovered over twenty-five years 
ago (Klebesadel, Strong, \& Olson 1973), and are now detected 
by BATSE at a rate of about one per day (Meegan et al. 1996). 
Their energy sources and emission mechanisms are unknown, and 
their distance scale remains a matter of controversy (Lamb 1995; 
Paczy\'nski 1995).  The solution of this longstanding mystery 
may finally be near with the recent detection of long-wavelength 
counterparts to GRBs (Costa et al. 1997a; van Paradijs et al. 1997;
Piro et al. 1997a).
 
On 1997~May~8.9 (UT) the Gamma-Ray Burst Monitor onboard the 
Italian-Dutch satellite BeppoSAX was triggered by a 15~s long 
transient event (Costa et al. 1997b) in the 40-700~keV energy 
range (peak flux $F_{\gamma}$ = (5.6 $\pm$ 0.7)$\times 10^{-7}$ 
erg s$^{-1}$ cm$^{-2}$), which was detected also by one of the 
BeppoSAX Wide Field Cameras, with an intensity equivalent to 1 
Crab unit in the 2-30 keV range (Costa et al. 1997b), by BATSE 
(Kouveliotou et al. 1997), and by Ulysses (Hurley 1997).
Within the $3^{\prime}$ error circle determined by the Wide Field 
Camera detection no other X-ray source was observed (Heise et al. 
1997).  Follow-up target-of-opportunity observations with the 
BeppoSAX Narrow Field Instruments in the 0.5-10 keV range on May 
9.1, 11.8, 13.1 and 15.1 revealed a previously unknown X-ray source 
which declined in intensity by a factor of four (Piro et al. 
1997b;c), supporting a possible identification of the transient 
as the X-ray counterpart of GRB970508.

An optical source was detected in the BeppoSAX Wide Field Camera 
error box with initial magnitude V$ = 21.5$ (Bond 1997).  The next
day, the source was seen to brighten by one magnitude.  Positional 
confirmation and refinement was obtained at Palomar (Djorgovski et
al. 1997) and from radio observations (Frail et al. 1997).  UBVRI 
photometry of the OT by other observers at many sites was taken 
during the following days.  In the R band, which is the best sampled, 
the flux was constant or slowly declining in the first 8 hours 
after the GRB, then it rose by a factor 5 in $\sim$40 hours.  
After a maximum on May 10.8, the brightness subsided monotonically 
with an approximately power-law temporal dependence 
$f(t) \propto (t - t_0)^{-p}$, where $t_0$ is the time of the GRB 
detection, with slope $p = 1.17 \pm 0.04$ (1$\sigma$), as determined 
through a weighted least-squares fit (reduced $\chi^2 \simeq 0.6$) 
to the decaying portion of the light curve (see Figure 1, and the 
caption for references).  This is reminiscent of the optical decline 
of GRB970228, which is generally well modeled by a temporal power-law 
of index $p = 1.1$ (Fruchter et al. 1997).

The spectrum of the OT between 3500 and 8000 \AA\ is well fit by 
a power-law $f_\nu \propto \nu^{-\alpha_\nu}$ with $\alpha_\nu 
\sim 1$ at all epochs for which simultaneous photometry is available 
in at least four filters, agreeing with spectrophotometry made at 
the Keck II telescope (Metzger et al. 1997a).  This corresponds 
to a flat $\nu f_\nu$ spectrum.  Keck spectroscopy also reveals 
absorption systems at $z = 0.767$ and 0.835 superposed on the 
continuum as well as [O II] line emission at $z=0.835$ (Metzger et 
al. 1997b).  The absence of Lyman-$\alpha$ absorption forest and of 
a continuum decrement suggests that its redshift, $z \simlt 2.3$.  
The presence of Mg I absorption and [O II] emission in the Keck 
spectrum suggest a line of sight through a dense interstellar medium; 
however, the only potential host galaxy detected from ground-based 
imaging is a faint blue object lying $5^{\prime\prime}.2$ away 
from the OT (Djorgovski et al. 1997). 

Here we present imaging using the newly deployed HST instruments 
STIS and NICMOS.  These observations were designed to search for a 
host galaxy and to obtain late-time photometry of the OT of GRB970508. 
Data reduction and analysis are described in \S 2, the results are 
reported in \S 3 and discussed in \S 4.

\section{Observations and data analysis}

Four 1250-second exposures were obtained of the GRB970508 field 
using the STIS CCD in Clear Filter mode during 1997 June 2.52-2.66 
(UT). In spite of the relatively short integration time, the high 
throughput and wide effective band-pass of the STIS CCD combined 
to make this one of the deepest images yet taken by HST -- only 
about one magnitude shallower than the Hubble Deep Field in the 
F606W filter of the HST Wide Field and Planetary Camera (Williams et
al. 1996).  The four exposures were dithered to allow removal of 
hot pixels and to obtain the highest possible resolution.  The images 
were bias subtracted, flat-fielded, corrected for dark current and 
calibrated by the newly created STIS pipeline.  The final image 
(Figure 2) was created and cleaned of cosmic rays using the 
Variable-Pixel Linear Reconstruction technique (Fruchter \& Hook 1997).  

Four exposures of 514 seconds each were also made with the NICMOS 
Camera 2 on 1997 June 2.67-2.74 (UT).  The four exposures were 
dithered using the NICMOS spiral dither pattern. The F160W filter 
(close to the standard near-infrared H band) was used as this filter 
covers the wavelength range where the HST background light is at a 
minimum.  With $0^{\prime\prime}.075$ pixels, NICMOS Camera 2 gives 
reasonably well-sampled diffraction-limited images at around 1.6 $\mu$m.  

At the time of the observations the understanding of the NICMOS 
instrument was in a very preliminary state, and as a result, during 
two of the four dither configurations the default positioning placed 
the OT on a detector column of very limited quantum efficiency.  
In addition, the observations were broken up by the scheduling program 
into two sessions, the second of which occurred immediately after 
the telescope had passed through the South Atlantic Anomaly, a region 
of particularly high particle background.  Therefore, two of the four 
images show significant noise caused by persistent charge left by 
cosmic rays.  These problems, combined with the early state of 
the calibration images has made the data less powerful, and our 
calibration less precise than nominal expectation.  Nonetheless, 
the OT point-like source is easily visible in all of the data sets, 
with a signal-to-noise ratio of about 15 in a single 500-second exposure.

\section{Results}

The photometric calibration of the images was done using the synthetic
photometry package SYNPHOT in IRAF/STSDAS. For STIS, given the broad-band 
response of the instrument, a power-law spectral shape with index 
$\alpha_\nu \simeq 1$ was assumed, based on Keck spectrophotometry and 
on the photometric colors. This yields for the OT V$=23.45 \pm 0.15$ 
($1\sigma$) and R$ = 23.10 \pm 0.15$.  Although the STIS photometric 
calibration is still preliminary, we find agreement with ground-based 
imaging of other stars in the field at a level of $0.1$ mag (Djorgovski
1997).  For NICMOS, our calibration gives an OT magnitude of H = $20.6 
\pm 0.3$.  However, as already noted, the NICMOS photometric calibration 
is still in a very preliminary state, and we have no corroborating 
ground-based data.  The large error we have assigned reflects the 
uncertainty of the photometric calibration.

In the near-infrared, the faint galaxies located at North-East (G1) and 
North-West (G2) can be detected in an image smoothed with a boxcar
of 4 pixels and are found to have fluxes of 0.8 $\pm$ 0.1 $\mu$Jy 
(H = 22.8) and 1.9 $\pm$ 0.1 $\mu$Jy (H = 21.9), respectively.  Since 
the STIS CCD response peaks in the V band, their R magnitudes have been 
interpolated assuming a power-law spectrum consistent with the V -- H 
color. The results are R = $24.8 \pm 0.2$ (G1) and R = $25.5 \pm 0.2$ 
(G2).  G1 is extremely blue, as reported by Djorgovski et al. (1997), 
and the colors are consistent with a rapidly star-forming galaxy at any 
reasonable redshift.  G2 is somewhat redder, but has the colors of a 
nearby late-type spiral galaxy whose spectrum has been shifted to 
$z \sim 0.7$.  Therefore, either of these objects could be responsible 
for the absorption seen in the Keck spectrum at $z = 0.767$.  As we 
explain below, we doubt that either of these galaxies could produce 
the absorption system at $z=0.835$, which appears to be caused by a 
dense interstellar medium in a low-ionization state.

The R band STIS magnitude lies within the $1\sigma$ uncertainty of the 
extrapolation to June 2.5 of a power-law decay fit to earlier data. An 
R band measurement (R = 23.4) taken at Keck after our HST observation 
(5 June) confirms this trend (see Figure 1).  Under the assumption that 
this decay law correctly reflects the behavior of the OT, the lack of 
any evidence of flattening in this temporal descent allows us to put a
$3 \sigma$ upper limit on an underlying galaxy of R=24.4.  To further 
constrain the luminosity of an underlying galaxy, we have subtracted a 
scaled stellar point spread function (PSF), obtained from an earlier 
STIS image of the globular cluster $\Omega$-Cen.  The residual, which 
is less than 10\% of the flux of the star, is consistent with that 
expected due to changes in telescope focus and the necessity of using 
a PSF star translated from the position of the OT on the detector.

We further tested our ability to detect nebulosity by creating an 
artificial ``compact galaxy'' by convolving the PSF with a Gaussian 
of intrinsic FWHM = $0^{\prime\prime}.15$.  Such an object would be 
just sufficiently extended to be distinguished from a point-like source 
in the Hubble Deep Field images.  After subtracting a PSF, residuals 
are easily visible in the image if the ``galaxy'' is 1-2 magnitudes 
fainter than the OT, or  R $\simeq 24-25$.  They become difficult 
to discern when the galaxy's magnitude is fainter than R $\simeq 26$.  
However, for R $>$ 25.5 one approaches a regime where a number of 
unresolved blue objects were found in the Hubble Deep Field: were one 
of those associated and well-aligned with the OT it would be subtracted 
out with the PSF.

As a final test we added the OT onto the image of G1 and again fit and 
subtracted a PSF. In this case the residuals were both immediately 
obvious to the eye and statistically significant.  We therefore believe 
that any underlying galaxy must be no brighter than R$=24.5$; if it is 
an extended object with a scale size greater than a few tenths of an 
arcsecond, it must be fainter still.  Thus, a source of the magnitude 
and shape of that reported for GRB970228 (Sahu et al. 1997a) would 
have been easily detected.

Similarly, after subtraction of a scaled artificial PSF, the NICMOS 
image is also consistent with sky noise statistics and we estimate 
any underlying, extended component must have H$>22.2$ within 
$0^{\prime\prime}.4$ of the point-like source.  Furthermore, if one 
uses ground-based I and K band measurements (Kopylov et al. 1997; Morris 
et al. 1997) obtained on May 13 to interpolate an H band magnitude and 
scale this value to the time of the NICMOS observations, according to 
a $p \simeq 1.2$ power-law decay, one predicts a flux of 
$3 \pm 1 ~ \mu$Jy, fainter than but in rough agreement with the NICMOS 
measurement of $6.2 \pm 1.5 ~ \mu$Jy.  This comparison is therefore 
consistent with our conclusion that a superposed galaxy could only
contribute a small fraction of the total observed H band flux.

\section {Discussion}
 
The absence of any obvious visible host galaxy is particularly striking 
given the detection of Mg I absorption and [O II] emission in the Keck 
spectra.  The Mg I indicates that the absorbing medium is not highly 
excited and the [O II] implies that active star formation is occurring.  
While there are several galaxies with V$> 24.5$ within a few arseconds 
of the OT, this corresponds to a projected distance of tens of kiloparsecs 
at $z=0.8$.  It seems unlikely that either the high density or low 
excitation necessary for the Mg I line could be maintained this far out 
in a galactic halo -- indeed of the 103 quasars in the compilation by 
Steidel and Sargent (1992), only 19 exhibit Mg I absorption lines, 
and none of these has rest frame line equivalent widths or ratios of Mg 
I to Mg II absorption as large as that seen in the spectrum of the OT 
of GRB970508.  Therefore, we believe that the absorbing medium 
responsible for these lines is presently hidden by the light from the 
OT and is almost certainly the underlying host galaxy. 

The implications of the redshift on the nature of the host are equally
profound.  Were the GRB at $z \sim 2$ (the upper limit suggested by the 
Keck spectrum of the OT) the $K$-correction and  cosmological dimming 
of the source would allow the host to be as bright as the knee of the 
galaxy luminosity function, $L^*$, and yet evade detection in both 
the STIS and NICMOS images.  On the other hand, if the OT is located 
at $z=0.8$, the host must be considerably fainter than $L^*$.  The 
[O II] emission superposed on the OT continuum suggests that the host 
is rapidly star-forming, and therefore quite blue (and thus would have 
a negligible $K$-correction at this redshift).  An apparent magnitude 
of R$=24.5$ then corresponds to an absolute blue magnitude of 
$-18.6 \pm 0.5$, where the error represents the uncertainty in the
cosmological parameters. 

One might be concerned, however, that we are not observing an optical 
source associated with GRB970508, but rather an extragalactic supernova 
or active galactic nucleus (AGN) located within the BeppoSAX positional 
error box.  However, the Keck spectrum indicates that such an interloper 
would have to be at a minimum redshift of $z = 0.8$, and thus we would 
be observing its restframe ultraviolet emission.  The absolute 
magnitude of a Type Ia supernova (the brightest type) in U band near 
maximum is $\sim~ -20$ (Leibundgut \& Tammann 1990). At a redshift of 
$z=0.8$, this would produce an apparent R band magnitude of $\sim 23.5$ 
or several magnitudes below the peak of the OT to GRB970508.  If a 
supernova were at a higher redshift, the discrepancy between predicted 
and observed magnitude would be even greater.  In addition, the spectrum 
of the OT shows none of the strong, broad features so prominent in 
Type Ia supernovae (Panagia 1987).  We therefore conclude that the OT 
is not an extragalactic supernova.  The suggestion that the OT is an 
AGN appears equally unlikely.  We know of no AGN that has exhibited 
behavior at all reminiscent of the regular, five magnitude power-law 
decline of this source.  Rather, AGNs are characterized by irregular, 
unpredictable optical variability (see reviews by Clavel 1994; Wagner
\& Witzel 1995; Ulrich, Maraschi, \& Urry 1997).

The upper limit to the host galaxy absolute magnitude,
${\rm M}_{\rm B} = -18.6$, estimated for a redshift of $z=0.8$, 
corresponds to about one tenth of $L^*$.  However it is now known that 
the star-formation rate in the universe was approximately a factor of 
ten higher at $z \sim 1$ (Lilly et al. 1995), and that the majority of 
this star formation probably occurred in galaxies less luminous than 
$L^*$ (see Babul \& Ferguson 1996, and references therein for a 
discussion of this question).  At least two models for the creation of 
GRBs -- the merging of neutron star binaries (Narayan, Paczy\'nski, \& 
Piran 1992) and failed supernovae (Woosley 1993) -- directly associate 
GRBs with the formation of massive stars.  In the latter case, the GRBs 
would generally occur at the sites of star formation, while in the case 
of merging neutron stars, the kick given to the binary at the birth of 
the second neutron star could cause the binary to travel up to tens of 
kiloparsecs during the $\simlt 100$ million year timescale of the merger 
(Tutukov \& Yungelson 1994).  

If formation of GRBs is indeed tied to star formation, the location of 
these objects in relation to their hosts should provide  strong clues to 
the nature of the birth mechanism.  Furthermore, the rate of GRBs with 
redshift would increase far more rapidly than the co-moving volume of 
the universe, for it would be proportional to the rate of star formation
multiplied by the co-moving volume.  The cosmological density of GRBs 
should then follow the general rise and fall of star formation with 
redshift (Madau et al. 1996).  Indeed, it appears possible that GRBs 
may eventually provide our best tool for measuring the star-formation 
history of the universe.

Although we may eventually find that GRB970508 does indeed lie in a 
faint, star-forming galaxy, at present the observations underline the 
longstanding issue of the lack of detected GRB host galaxies (Fenimore 
et al. 1993; Schaefer 1994; though see Larson 1997) and point to the 
necessity of pursuing deep and systematic studies of GRB environments.  
This will only be achieved through the dedication of significant observing
time on HST as well as on large, ground-based optical telescopes.

\acknowledgements
 
We thank Bob Williams  for allocating Director's Discretionary time 
to this program.  We would also like to express gratitude to S. Baum, 
I. Busko, J. Christensen, H. Ferguson, J. Hayes, E. Huizinga, A. Roman 
and P. Stanley for their help in observation planning, and assistance 
with data reduction.  We acknowledge the major efforts of the STIS and 
NICMOS Investigation Definition Teams, in building these powerful new 
instruments, and providing useful calibrations at such an early stage, 
and particularly thank R. Thompson and  B. Woodgate.  We benefitted 
from discussion with P. Caraveo, G. Djorgovski, D. Frail, G. Ghisellini, 
J. Gorosabel, P. Madau, K. Noll, M. O'Dowd, N. Panagia, R. Scarpa, C. 
Steidel, and  C. M. Urry.  C. Alcock, D. Dal Fiume, A. Guarnieri, J. 
Heise, L. Metcalfe are acknowledged for their support of this project.  
M. Feroci, L. Nicastro, E. Palazzi, and E. Pian acknowledge 
financial support from the Italian Space Agency (ASI).

\noindent
{\bf Authors' note}:  The reduced HST data discussed in this paper are 
available from http://www.stsci.edu/$\sim$fruchter/GRB/data\_970508.
 
\newpage
 

 
\newpage

\cl{\bf Figure Captions}

\bigskip

\figcaption{Light curve of the OT of GRB970508.  Photometry is taken 
from Pedersen et al. (1998), Castro-Tirado et al. (1997),  Sahu et al.
(1997b, the R magnitudes have been interpolated from the V and I 
magnitudes assuming a power-law spectrum), Djorgovski et al. (1997), 
Galama et al. (1997), Schaefer et al. (1997), Kopylov et al. (1997), 
Mignoli et al. (1997), Groot et al. (1997), Garcia et al. (1997), 
Chevalier \& Ilovaiski (1997), Metzger et al. (1997b) (open circles)
and HST-STIS (filled circle). All magnitudes have been converted to 
Kron-Cousins R. Uncertainties have been rounded up to 0.1 magnitudes
when smaller values were reported in the literature, to take into
account possible systematic photometric offsets due to instrumental 
differences.  The power-law fit to the 10.5-22 May data (solid line) 
is reported along with the 1$\sigma$ uncertainty  range (dashed lines).}

\figcaption{The HST-STIS image of the field surrounding the OT of 
GRB970508.  A region of $22^{\prime\prime}.5 \times 16^{\prime\prime}.25$ 
is shown.  The arrows indicate the North and East directions. The two 
galaxies at $\sim 5^{\prime\prime}$ distance from the OT, referred
to in the text as G1 and G2, are on the upper left and upper right,
respectively.  The final frame was created using $0^{\prime\prime}.025$ 
pixels, one-half the size, in linear dimension, of the original STIS 
pixels. The image has a FWHM of $0^{\prime\prime}.09$ and a limiting 
$10 \sigma$ sensitivity in a $0^{\prime\prime}.5 \times 0^{\prime\prime}.5$ 
box of V$=27.4$.  The HST-NICMOS image shows only a moderate 
signal-to-noise point-like source at the position of the OT, and has 
therefore not been displayed.}
 
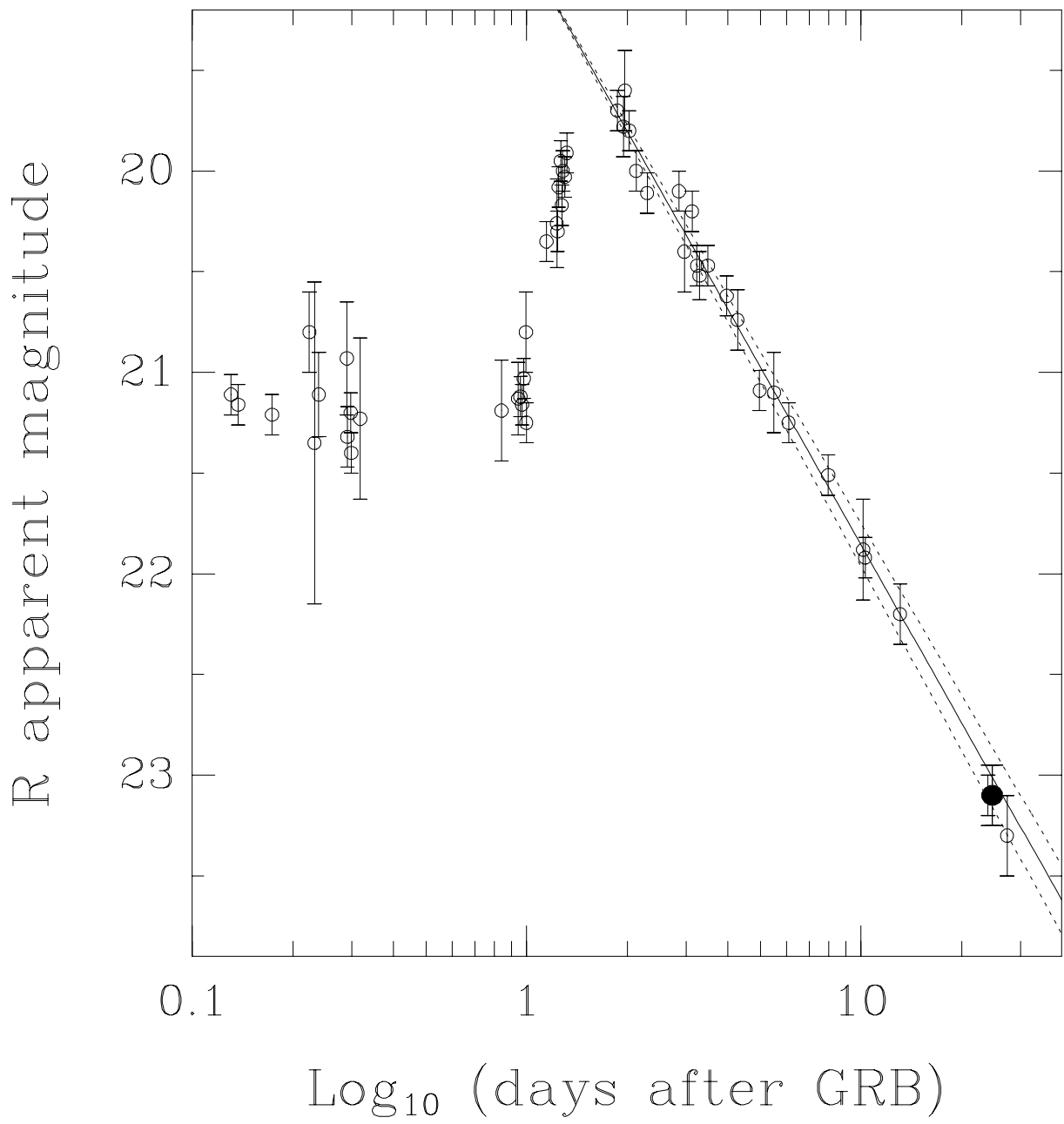
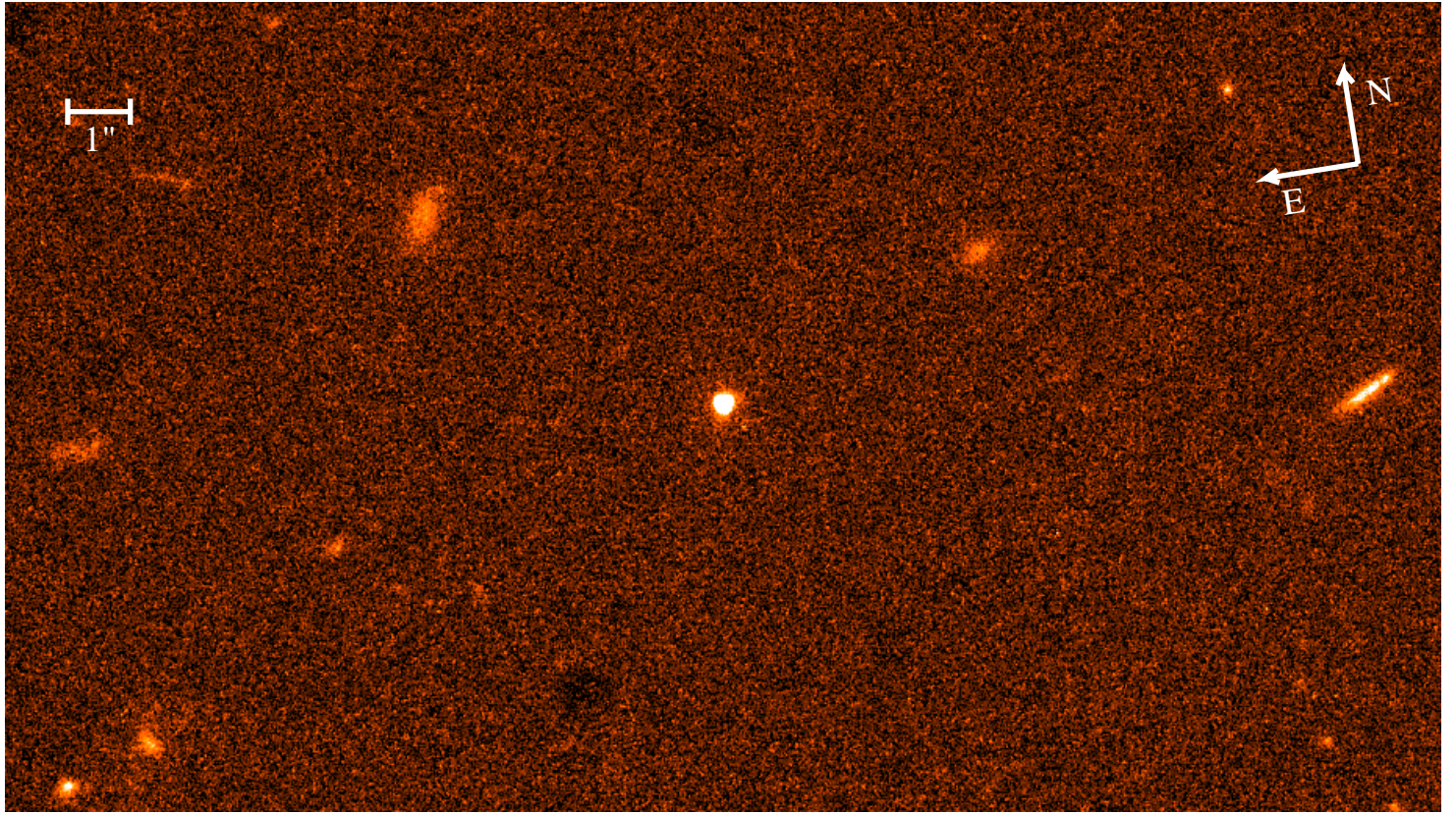
\end{document}